\documentclass[aps,prl,twocolumn,groupedaddress]{revtex4}

\usepackage{graphicx}

\begin{document}


\title{Direct calculation of the crystal-melt interfacial free energies 
for continuous potentials: Application to the Lennard-Jones system}


\author{Ruslan L. Davidchack}
\affiliation{Department of Mathematics and Computer Science, University
of Leicester, Leicester LE1 7RH, UK}
\author{Brian B. Laird}
\affiliation{Department of Chemistry, University
of Kansas, Lawrence, Kansas 66045, USA}


\date{\today}

\begin{abstract}
Extending to continuous potentials a cleaving wall molecular-dynamics simulation method
recently developed for the hard-sphere system [Phys.Rev.Lett {\bf 85}, 4751 (2000)], we 
calculate the crystal-melt interfacial free energies, $\gamma$, for a 
Lennard-Jones system as functions of both crystal orientation and temperature.
At the triple point, $T^* = 0.617$, the results are consistent with an earlier 
cleaving potential calculation by Broughton and Gilmer [J. Chem. Phys. {\bf 84},
5759 (1986)], however, the greater precision of the current calculation allows us
to accurately determine the anisotropy of $\gamma$. From our data we find that, 
at all temperatures studied, $\gamma_{111} < \gamma_{110} < \gamma_{100}$. 
Comparison is made to the results from our previous hard-sphere  calculation and to recent 
results for Ni by Asta, Hoyt and Karma [Phys. Rev. B, {\bf 66} 100101(R) (2002)]. 
\end{abstract}


\maketitle


\section{Introduction}

The magnitude and orientational dependence (anisotropy) of the solid-liquid interfacial 
free energy, $\gamma_{sl}$,  is a primary controlling parameter in the kinetics and 
morpholgy of crystal growth from the melt\cite{Tiller91}, especially in the case of dendritic 
growth\cite{Boettinger00}. As a consequence, the ability to accurately measure or predict this 
quantity for specific materials is of significant technological and scientific importance. 
For most materials, the only experimental data for $\gamma_{sl}$ is extracted indirectly from 
nucleation data (assuming some level of classical nucleation 
theory)\cite{Tiller91,Woodruff73,Turnbull50}. Such indirect measurements tend to underestimate the 
actual interfacial free energy by 10-20\% and represent orientatial averages, so all information as
to interfacial anisotropy is lost. Direct experimental measurements, usually involving contact angle
studies, are quite difficult and relatively few in number\cite{Howe97} and, with the exception 
of a small number of studies on transparent organic materials\cite{Glicksman89,Muschol92}, are
not of sufficient precision to resolve anisotropy. This paucity of reliable direct experimental 
measurements on technologically useful materials (such as metals) has motivated the development
of a variety of novel computational methods to determine $\gamma_{sl}$ via molecular 
simulation\cite{Broughton86c,Davidchack00,Hoyt01,Morris02}. 

The interfacial free energy of a crystal-melt interface is defined\cite{Tiller91} as the 
reversible work required to form a unit area of interface.  In a simulation this can be accomplished
by constructing a continuous thermodynamic path between an initial system consisting of separated
bulk crystal and liquid to a final state containing an interface. The value of $\gamma_{sl}$ is then 
determined by thermodynamic integration\cite{Frenkel02} along that path. This is a tedious process 
and care must to be taken to ensure that the process is reversible, i.e., intergration along the
path in both the forward and backward directions yields the same result (no hysteresis).  
The first such calculation on a crystal-melt interface was performed by Broughton 
and Gilmer\cite{Broughton86c} on a system of particles interacting with a LJ potential (truncated so 
that both the potential and the force vanish at $2.5\sigma$, where $\sigma$ is the usual LJ diameter).
To perform the thermodynamic integration they employed external cleaving potentials that were slowly 
turned on to separate the samples. The precise (rather complex) functional forms of the cleaving potentials
were chosen, more or less by trial and error, to minimize hysteresis.  The values of $\gamma_{sl}$ were
determined to be 0.35(2), 0.34(2) and 0.36(2) (in units of $\epsilon/\sigma^2$) for the [111],[100],
and [110] crystal orientations, respectively. The numbers in parentheses give the uncertainties in 
the last digit shown. The main source of this error is the small amount of hysteresis 
in the cleaving of the liquid phase.  Unfortunatly, the precision of these results was not sufficient 
to resolve the anisotropy of the interfacial free energy for this system.  

Recently, we have developed a modification of the Broughton
and Gilmer approach in which planar cleaving {\it walls}, as opposed to cleaving {\it potentials}, are 
used to separate the phases. These walls are constructed out of the same type of particles as 
present in the system, with a 2-d geometry consistent with the symmetry of the interfacial orientation
under study. This method was originally applied to the crystal-melt interface of a system of hard 
spheres\cite{Davidchack00} and was shown to have sufficient precision to resolve the anisotropy with [111]
being the lowest free energy face, followed by [100] and [110].  The cleaving wall method is complementary 
to a method due to Hoyt, Asta and Karma\cite{Hoyt01} in which the interfacial free energy is extracted from 
the interfacial stiffness, which is determined from the spectrum of fluctutations in interfacial position.
The cleaving-wall method has an advantage in that it requires simulations with an order of 
magnitude fewer particles than those required for the fluctuation method (${\cal O}[10^4]$ vs. 
${\cal O}[10^5]$). The precision in the raw values of the interfacial free energies is greater for the
the cleaving wall method than in the fluctuation approach (even considering the far smaller system samples); 
however, the fluctuation method yield somewhat more precise values of the anisotropy parameters since these
are obtained from the anisotropy of the interfacial stiffness, which is far more anisotropic than the 
interfacial free energy. 

In this work, we extend our cleaving wall approach to systems of particles interacting with 
continuous potentials, specifically applying it to the system of truncated Lennard-Jones (LJ) particles 
considered by Broughton and Gilmer. Our results at the
triple point are consistent with Broughton and Gilmer's calculation, but are of greater precision 
allowing us to resolve the anisotropy, which is found to differ slightly from that for the hard-sphere
system\cite{Davidchack00}. In addition, we determine the temperature dependence of $\gamma_{sl}$ along
the coexistence curve. The magnitude of $\gamma_{sl}$ is shown to scale roughly linearly with the melting 
temperature (as predicted by the hard-sphere model\cite{Laird01}). 

\section{The Cleaving Wall Method}

\begin{figure}
\includegraphics[bb=70 230 450 700,width=5cm]{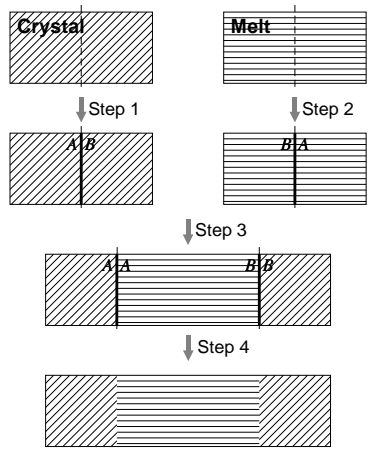}
\caption{ \small Illustration of the four steps
reversible process of creating the crystal-liquid interface from
separate bulk systems.  Dashed lines show the location of the
cleaving planes and thick solid lines represent the cleaving potental.
Letters $A$ and $B$ label the crystal and liquid on either side of 
the cleaving planes for better illustration of the boundary 
conditions rearrangement in step 3.  Periodic boundary conditions 
are assumed on all boundaries of the simulation boxes.}
\label{fig1}
\end{figure}

The direct determination of the excess free energy of the crystal-fluid interface of a model
system within a computer simulation can be achieved by thermodynamic integration along a 
reversible path beginning with separate crystal and fluid bulk systems prepared at the coexistence 
conditions and ending with a system containing a crystal-liquid interface at equilibrium with the 
surrounding bulk phases. Construction of such a path requires the development of a procedure to 
reversibly cleave a simulation box into two non-interacting systems.  Following the prescription of 
Broughton and Gilmer\cite{Broughton86c}, we identify the following steps in the process of creating the
crystal-liquid interface (see Fig.~\ref{fig1}):
\begin{description}
\item[Step 1:] Split the crystal bulk system with a suitably chosen
``cleaving'' potential while maintaing the periodic boundary conditions
\item[Step 2:] Split the liquid system in a similar way
\item[Step 3:] Juxtapose the cleaved crystal and liquid systems by
rearranging the boundary conditions while maintaining the cleaving
potentials
\item[Step 4:] Remove the cleaving potentials from the combined 
system.
\end{description} 

The interfacial free energy is calculated as the total work required to
perform the above steps divided by the area of the constructed interface.
In addition to the coexistence conditions, the result will also
depend on the orientation of the crystal with respect to the 
interfacial plane. We will refer to the plane along which the crystal and 
liquid systems are split as the {\em cleaving plane}.  The location of the
cleaving plane in the crystal system is chosen in the center of the simulation box 
between two crystal layers, while in the liquid system the precise location is 
arbitrary.  

The reversibility and precision of the thermodynamics integration
process are very sensitive to the choice of cleaving potentials.
The requirements for the cleaving potential are two-fold:  First,
the cleaving potential should perturb the system as little as possible.
As a consequence, it is desirable that, in step 2, the potential 
introduces structure into the cleaved liquid that is compatible with 
the structure of the crystal layers.  Second, the cleaving potential 
must be strong enough to prevent the particles from crossing the 
cleaving plane.  Otherwise, the rearrangement of the boundary 
conditions in step 3 cannot be performed.

Broughton and Gilmer\cite{Broughton86b} carefully designed a set of
cleaving potentials for the truncated Lennard-Jones (LJ) system.  However, 
their approach for constructing the cleaving potential was 
optimized specifically for the LJ  system and is not easily adaptable to a general case.
Here we outline an approach that is easily adaptable to systems with different 
interparticle interaction potentials and different crystal structures.

As was demonstrated in our recent calculation of the interfacial free energy for the hard-sphere
system\cite{Davidchack00}, the appropriate structure
in the interfacial region of the fluid can be easily introduced by
the potential of a pair of ``cleaving walls'', made of properly 
oriented crystal layers, each interacting only with the 
particles on the opposite side of the cleaving plane. 
The wall particles are held fixed at the crystal lattice sites.
When the two walls are far away from the cleaving plane, they do not 
interact with the system.  Moving the walls closer to the cleaving 
plane introduces a cleaving potential whose structure is similar to
that outside the crystal layers of the studied system.
When a liquid system interacts with such a cleaving wall, it is
expected to form an interfacial layer structure similar to that at
the crystal-liquid interface.  To achieve this, the
interaction potential of the wall particles must be similar to that 
of the system particles.  

Since interactions in the crystal are usually dominated by the 
short-range repulsive part of the potential, it is sufficient to choose 
the interaction potential of the wall particles as a monotonically
decreasing function $\phi(r)$ with a relatively small cut-off radius
$r_w$, which models the repulsive core of the interparticle potential
in the system under investigation.  Interaction of the system with 
each cleaving wall is then given by
\begin{eqnarray}
 \Phi_{1}({\bf r}; z) &\!\!=\!\!& 
\sum_j\phi(|{\bf r}-{\bf R}_j^{(1)}+{\bf n}z|)\,,\nonumber\\
 \Phi_{2}({\bf r}; z) &\!\!=\!\!& 
\sum_j\phi(|{\bf r}-{\bf R}_j^{(2)}-{\bf n}z|)\,,
\label{eq:u12walls} \end{eqnarray}
where ${\bf R}_j^{(1,2)}$ are the positions of the particles forming 
the walls, ${\bf n}$ is a unit vector normal to the cleaving plane,
and $z$ measures the distance of the walls to the cleaving plane.   

Next, we need to ensure that the system on the one side of the cleaving
plane interacts only with the wall on the other side of the plane.
To achieve this, we use the monotonic character of the potential
$\phi(r)$ and define the cleaving potential as the {\em minimum} of 
the two wall potentials, namely
\begin{equation}
  \Phi({\bf r}; z) = \min(\Phi_1, \Phi_2)\,,
\label{eq:clvmin} \end{equation}
which decays to zero away from the cleaving plane, as long as
the potential $\phi(r)$ is chosen to be a positive monotonically
decreasing function.  To remove discontinuity of the gradient
of $\Phi$ at the points where $\Phi_1 = \Phi_2$, we slightly modify the
minimum function as follows:
\begin{equation}
  m(x,y) = \left\{ \begin{array}{lc} x, & x \le y - \delta(x+y)\\
  y, & x \ge y + \delta(x+y)\\ p(x,y), & \mbox{otherwise} \end{array} 
  \right.
\label{eq:mxy} \end{equation}
with
\begin{equation}
  p(x,y) = \frac{x+y}{4}(2 - \delta) - \frac{(x-y)^2}{4\delta(x+y)}\,,
\label{eq:pxy} \end{equation}
and parameter $\delta$ characterising the relative width of the 
interpolation region.  We set $\delta = 0.25$ for the present study.
The cleaving potential is defined as
\begin{equation}
  \Phi({\bf r}; z) = m(\Phi_1, \Phi_2)\,.
\label{eq:clvm} \end{equation}

Even though the crystal system does not require additional ordering and,
as proposed by Broughton and Gilmer, can be cleaved with a short-range
repulsive potential centered at the cleaving plane, we have found that
using the same cleaving potential for both crystal and liquid systems
results minimizes the error during the thermodynamic integration in
step 3.

To calculate the reversible work in steps 1, 2, and 4, we can use
the wall position $z$ as the integration coordinate.  The reversible
work is thus determined by evaluating the integral
\begin{equation}
  w_{1,2,4} = \int_{z_i}^{z_f}\! \left\langle 
  \frac{\partial\Phi}{\partial z}\right\rangle dz\,,
\label{eq:work124} \end{equation}
where the angle brackets denote averaging over a simulation run at a
fixed cleaving wall position.  In steps 1 and 2, the initial position
of the cleaving walls, $z_i$, is just outside the range of the 
interaction potential determined by the cut-off radius $r_w$.
The final wall position $z_f$ is determined by the requirement that
the cleaving potential is sufficiently strong to prevent the particles
from crossing the cleaving plane.  In step 4, the initial and final 
positions of the walls are reversed.  Because of the repulsive 
character of the cleaving potential, the work in steps 1 and 2 is 
expected to be positive, while in step 4 it is negative.

In step 3, the boundary conditions are gradually rearranged using a
coupling parameter $\lambda$.  The total interaction energy in step 3
is given by
\begin{eqnarray}
  U(\lambda)& =& (1-\lambda)\sum_{i<j,\,AB} u(r_{ij})  \nonumber \\
&& + \lambda \sum_{i<j,\,AA}u(r_{ij}) 
+ \sum_i \Phi({\bf r}_i; z_f)\,,
\label{eq:step3pot} 
\end{eqnarray} 
where the letters $(AB)$ refer to the boundary conditions with 
crystal-crystal and liquid-liquid interactions across the 
cleaving planes, and $(AA)$ refer to the boundary conditions with 
crystal-liquid interactions (see the diagram).  The last term 
represents the cleaving potential at the final wall positions in 
steps 1 and 2.  The work done during step 3 is determined from 
the integral
\begin{equation}
w_3 = \int_0^1\!\left\langle \frac{\partial U}{\partial \lambda}
\right\rangle d\lambda\,, 
\label{eq:work3} \end{equation}
where
\begin{equation}
  \frac{\partial U}{\partial \lambda} = \sum_{i<j,\,AA} u(r_{ij}) - 
  \sum_{i<j,\,AB} u(r_{ij})\,.
\label{eq:dudl} \end{equation}

\section{Results for truncated Lennard-Jones potential}

In order to have direct comparison with the results of
Broughton and Gilmer\cite{Broughton86b}, we have used the same 
modification of the Lennard-Jones potential, namely \cite{Broughton83}
\[
  u(r) = \left\{\begin{array}{ll} 4\epsilon\left[\left(
  \frac{\sigma}{r}\right)^{12}-\left(\frac{\sigma}{r}\right)^6\right]
  + C_1 \quad &r \le 2.3\sigma \\ 
  C_2\left(\frac{\sigma}{r}\right)^{12} +
  C_3\left(\frac{\sigma}{r}\right)^6 &\\ \; \; + 
  C_4\left(\frac{r}{\sigma}\right)^2 + C_5\,, \quad
 & 2.3\sigma < r < 2.5\sigma \\ 0\,, \qquad \qquad & 2.5\sigma \le r\,.
  \end{array} \right.
\]
where $C_1 = 0.016132\epsilon$, $C_2 = 3136.6\epsilon$, 
$C_3 = -68.069\epsilon$, $C_4 = -0.083312\epsilon$, and 
$C_5 = 0.74689\epsilon$. (Note: the sign of $C_4$ was incorrectly reported as
positive in Broughton and Gilmer's original publication). 
This potential has a continuous first derivative and a small 
discontinuity in its second derivative at $r = 2.3\sigma$.

To achieve optimal performance of the simulation at all stages of the
cleaving process, we use a cell-assisted force evaluation method\cite{Rappaport95}.  
The simulation region is subdivided
into cells with edge sizes just exceeding the interatomic interaction 
range.  When atoms are assigned to cells according to their current 
position, only interactions between atoms in the same cell or in 
immediately adjacent cells needs to be considered.  Such a method is 
particularly useful in step 3, where interaction forces must be 
computed for two different boundary conditions 
[see Eq.~(\ref{eq:step3pot})].  With the cell-assisted method the
additional computational effort is limited to the cell layers 
immediately adjacent to the cleaving planes. (Note, the cell subdivision is carried out 
in such a way that the cell boundary coincides with the cleaving plane.)

The insertion of the exteral cleaving potentials ("walls") breaks the translational symmetry of the 
system Hamiltonian, linear momentum is no longer strictly conserved. This leads to a problem for long runs
in that the crystal, taken as a whole, can drift relative to the cleaving plane. To prevent this, we 
immobilize the innermost 2 crystal layers by assigning them an infinite mass. In the data-collection
runs these fixed planes were 20 crystal layers away from the nearest crystal-melt interface, and should 
not have a significant effect on the interfacial properties. To check this, we have performed smaller 
simulations with identical cross-sectional area, but fewer crystal layers (where the interface is about 12-14
crystal layers distant from the fixed planes) and obtain results identical, within simulation error, to
the larger systems. 

For the present study, we calculate $\gamma$ at both the triple point 
temperature of $T^* \equiv k_B T/\epsilon = 0.617$ (as determined by Broughton
and Gilmer \cite{Broughton86a}), and at higher temperatures
($k_B T/\epsilon = 1.0$ and  1.5).  The 
crystal-liquid coexistence parameters at these temperatures are 
summarized in Table \ref{tab:coex}. (Note the slightly 
negative pressure at $k_B T/\epsilon = 0.617$.  We have found that 
$k_B T/\epsilon = 0.618$ is a better estimate of the triple point 
temperature for the modified LJ potential used here.
However, to have a direct comparison with the results of 
\cite{Broughton86a}, we have calculated interfacial free energy
at the lower temperature.)  The parameters are obtained by 
monitoring long simulation runs of the crystal-liquid interfacial 
systems.  If the initial conditions are shifted from those at 
coexistence, such a system equlibrates itself through melting/freezing 
at the interfaces.  During the runs, the pressure tensor 
profiles \cite{Davidchack98} are monitored and the simulation 
box rescaled, if necessary, to remove any stress in the bulk crystal.
\begin{table}[h]
\caption{\label{tab:coex} \small Coexistence conditions for the modified
Lennard-Jones potential}
  \begin{ruledtabular}
  \begin{tabular}{|lrrr|}
  $T$, $k_B \epsilon^{-1}$ & $\rho_c$, $\sigma^{-3}$ & 
  $\rho_l$, $\sigma^{-3}$ & $P$, $\epsilon\sigma^{-3}$ \\ \hline
  0.617 & 0.945 & 0.828 & -0.02 \\  1.0 & 1.005 & 0.923 & 4.95 \\ 
  1.5 & 1.074 & 1.003 & 12.9
  \end{tabular}
  \end{ruledtabular}
\end{table}

\begin{figure}
 \includegraphics[width=5cm]{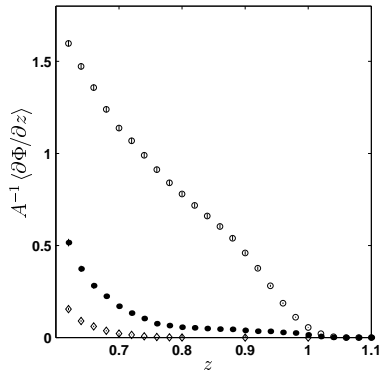}
\caption{
Integrand (Eq. 6) for thermodynamic integration in steps 1,2 and 4 for the [111] interface
at $T^*$ = 0.617.  The error bars are two standard deviations  plus estimated hysteresis error, where
appropriate.
}
\label{fig2}
\end{figure}

\begin{figure}
 \includegraphics[width=5cm]{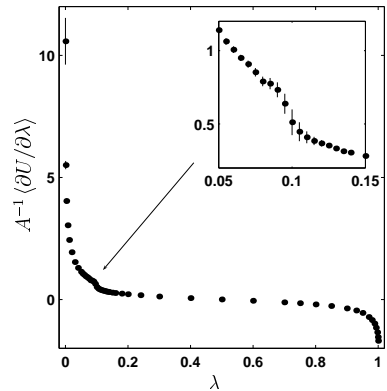}
 \caption{
Integrand for thermodynamic integration in step 3 for the [111] interface at
$T^*$ = 0.617. The inset shows a magnification of the region primarily affected by hysteresis, 
reflected by the larger than average error bars.  The error bars are two standard 
deviations  plus estimated hysteresis error, where appropriate.
}
\label{fig3}
 \end{figure}

At each of the three temperatures, we calculate the interfacial free
energy for three crystal orientations: [100], [110], and [111].
The cleaving wall is constructed from a single crystal layer for
the [100] and [111] orientations and from two layers for the [110] 
orientation (this was necessary to prevent atoms from crossing through this
non-close-packed face).  As discussed in the previous section, the cleaving 
potential $\phi(r)$ is constructed from the repulsive core of the LJ potential:
\begin{equation}
  \phi(r) = \left\{ \begin{array}{ll} 4\epsilon\left[\left(
  \frac{\sigma}{r}\right)^{12}-\left(\frac{\sigma}{r}\right)^6\right]
  + \epsilon\,, \quad&r < r_w = 2^{1/6}\sigma \\
  0\,, \qquad \qquad& r_w \le r\,.  \end{array}\right.
\label{eq:clvphi} \end{equation}
The results are summarized in Table \ref{tab:gamma}. 
For illustration of the method we show in 
Fig~\ref{fig2} the thermodynamic integration integrand (Eq. 6) for steps 1,2 and 4 for the [111] interface 
at $T^* = 0.617$. For that same interface and temperature, the integrand for step 3 (Eq. 7) is shown in 
Fig.~\ref{fig3} with an inset highlighting the region of maximum hysteresis error. 
\begin{table}[h]
\caption{\small Interfacial free energy (in units of $\epsilon\sigma^{-2}$) for the 
truncated Lennard-Jones potential for selected temperatures and crystal 
orientations.  Numbers in parentheses indicate two standard deviation error bars (plus an estimate
of the hysteresis error on the last digit(s) shown.}
\begin{ruledtabular}
  \begin{tabular}{|c|ccc|}
  & $T^*$ = 0.617 & 1.0 & 1.5 \\\hline 
  $\gamma_{100}$ & 0.371(3) & 0.562(6) & 0.84(2) \\
  $\gamma_{110}$ & 0.360(3) & 0.543(6) & 0.82(2) \\
  $\gamma_{111}$ & 0.347(3) & 0.508(8) & 0.75(3) 
  \end{tabular}
  \end{ruledtabular}
\label{tab:gamma}
\end{table}

Even though the relative statistical accuracy that can be achieved for a given 
duration of the simulation run is about the same for all temperatures and 
orientations (approximately 0.5\%), we see that the relative error range in 
Table~\ref{tab:gamma} increases with the temperature, especially for the [111] 
orientation. This increase is due to the observation that reversibility 
(as measured by the lack of hysteresis) of the thermodynamic integration process 
for the LJ system is more difficult to achieve for higher temperatures.

There are two sources of potential hysteresis in the four-step process of 
creating an interface. The first one is associated with the liquid ordering 
transition that occurs either at the end of step 2, or at the beginning of 
step 3. For lower temperatures, the hysteresis can be essentially eliminated 
by increasing the duration of the equilibration runs near the point of 
transition. However, the hysteresis for higher temperatures is more persistent, 
especially in the case of the [111] orientation. This may be due to a particular 
choice of the cleaving potential. More research is necessary to elucidate the 
origin of the hysteresis and ways to eliminate it.

The second source of hysteresis is the fluctuation in the interface 
position at the end of step 4. When the cleaving potential is removed from the 
interfacial system in step 4, the position of the interface is no longer tied to 
the cleaving plane. Because the system contains two interfaces, they can change 
their position without disturbing system equilibrium by the process of melting at 
one interface and simultaneous freezing at the other interface. The difficulty in 
verifying the reversibility of step 4 is that the mobility of the interfaces 
causes the reverse process to follow a slightly different thermodynamic 
integration path.  To deal with this problem, we try the reverse step 4 process 
on the interfacial systems after several equlibration runs of various duration 
and select the one with the path closest to the forward process.  The difference 
in the calculated work during the forward and reverse processes is accounted for 
in the estimate of the error range given in Table~\ref{tab:gamma}.

\section{Analysis and Summary}

The error bars on the calculations described above are small enough to resolve the anisotropy 
in the interfacial free energy for the Lennard-Jones system. While we have determined 
$\gamma_{sl}$ only for the [100], [110] and [111] directions, it is possibile to extract 
from these data some information as to the full angular dependence of the free energy. 
Defining the orientation unit vector $\hat{\bf n}$ as the unit vector perpendicular to the interfacial
plance, one defines an orientation dependent interfacial free energy $\gamma(\hat{\bf n})$, which 
can be parameterized by an expansion in terms of cubic harmonics. One such expansion, due to Fehlner and
Vosko, has been recently applied to the interfacial free energy of Ni/Cu alloys by Asta, 
et al.\cite{Asta02} In terms of the Cartesian components of  $\hat{\bf n} = \{n_1,n_2,n_3\}$, this 
expansion (truncated at sixth-order) is 
\begin{eqnarray}
\gamma(\hat{\bf n})& =& \gamma_0 \left [ 1 + \epsilon_1 \left ( \sum_{i=1}^3 n_i^4 - \frac{3}{5} \right )
\right .
\\
&& \left. + \epsilon_2 \left ( \sum_{i=1}^3 n_i^4 + 66 n_1^2n_2^2n_3^2 - \frac{17}{7} \right ) \right ] \;,
\nonumber
\end{eqnarray}
where $\gamma_0$ is the orientationlly averaged interfacial free energy and $\epsilon_1$ and $\epsilon_2$
are expansion coefficients. This parameterization has an advantage over the so-called 
"Kubic Harmonic" expansion\cite{Altman65} that has been used recently to parameterize free energy
anisotropy in metals\cite{Hoyt01,Morris02} in that the expansion terms are orthogonal\cite{Asta02}. 
For the orientations studied here this expansion becomes
\begin{eqnarray}
\gamma_{100}& =& \gamma_0 [1 + \frac{2}{5}\epsilon_1 + \frac{4}{7}\epsilon_2] \nonumber \\
\gamma_{110}& =& \gamma_0 [1 - \frac{1}{10}\epsilon_1 - \frac{13}{14}\epsilon_2] \nonumber \\
\gamma_{111}& =& \gamma_0 [1 - \frac{4}{15}\epsilon_1 + \frac{64}{63}\epsilon_2] 
\end{eqnarray}
Using our values for $\gamma$ for the [100], [110] and [111] orientations one can solve for the three 
parameters $\gamma_0$, $\epsilon_1$ and $\epsilon_2$:
\begin{eqnarray}
\gamma_0& =& \frac{10 \gamma_{100} + 16 \gamma_{110} + 9 \gamma_{111}}{35}\\ \nonumber
\epsilon_1& =& \frac{35\gamma_{100} - 8\gamma_{110} - 27\gamma_{111}}{22\gamma_0}\\ \nonumber
\epsilon_2& =& \frac{3}{22\gamma_0}(\gamma_{100} - 4\gamma_{110} + 3\gamma_{111})
\end{eqnarray}

The anisotropy parameters for the Lennard-Jones system studied here, as well as those for
the hard-sphere system calculated from the data from our earlier calculation\cite{Davidchack00}, are 
summarized in Table~\ref{tab:aniso}. Also included in Table~\ref{tab:aniso} are the values of 
$\frac{\gamma_{100}-\gamma_{110}}{\gamma_0}$ and $\frac{\gamma_{100}-\gamma_{111}}{\gamma_0}$, 
which also serve to quantify the anisotropy. The error bars in $\gamma_0$ are smaller than those for the 
raw interfacial free energies since $\gamma_0$ represents a weighted average of similar numbers - a process
that decreases statistical error. The anisotropy parameters represent differential quantities involving 
diffferences between similar numbers (which magnifies relative error) so the relative error bars for 
those quantities are larger than in the raw data. This is in contrast to the fluctuation  
method\cite{Asta02} where the relative error in the interfacial free energy is larger than in the current
study (despite the much larger systems), but the anisotropy parameters are more precisely 
determined due to the fact that in that method the anisotropy is determined from the 
anisotropy in the interfacial stiffness, which is far greater than that of the interfacial 
free energy itself. 
\begin{table}[h]
  \caption{\small Interfacial free energy anisotropy parameters for
the truncated Lennard-Jones at three different temperatures. For reference, the 
same quantities for the hard-sphere system [Reference 9] are also included.
Numbers in parentheses reflect 2-sigma error bars in the last digit(s). The units for $\gamma_0$ are
$\epsilon/\sigma^2$ for the LJ data and $kT/\sigma^2$ for hard spheres.}
\begin{ruledtabular}
\begin{tabular}{|c|ccc|c|}
& \multicolumn{3}{c|}{LJ (this work)} & HS $^9$\\
& $T^*$ = 0.617 & 1.0 & 1.5 & \\
\hline
$\gamma_0$ & 0.360(2) & 0.539(4) & 
0.808(13) & 0.617 \\
$\epsilon_1$  & 0.093(17) & 0.13(3) & 0.15(6) &0.07(3)\\
$\epsilon_2$  & -0.011(4) & -0.022(9) & -0.03(2) & -0.044(12)\\
$\frac{\gamma_{100}-\gamma_{110}}{\gamma_0}$ & 0.03(1) & 0.035(15) & 0.025(3) & -0.032(22) \\
$\frac{\gamma_{100}-\gamma_{111}}{\gamma_0}$ & 0.07(1) & 0.10(2) & 0.11(3)& 0.065(22) \\ 
\end{tabular}
  \end{ruledtabular}
\label{tab:aniso}
\end{table}
From the anisotropy parameters given in Table~\ref{tab:aniso} the trend is observed that the anisotropy
parameters increase in magnitude with increasing temperature, with $\epsilon_1$ becoming more
positive and $\epsilon_2$ becoming more negative. In comparison with the hard-sphere values, the value 
of $\epsilon_2$ approaches the hard-sphere value as $T$ gets larger, but the value of $\epsilon_1$, which
is at $T^*=0.617$ quite close to the hard-sphere value, diverges away from the hard-sphere
value as $T$ increases. Of course, in the limit of high temperatures, the Lennard-Jones system 
approaches an inverse twelth-power repulsive potential, not the hard-sphere potential, so one would expect 
the high temperature behavior of the anisotropy to approach that of the former potential. This implies that
any perturbation theory for the interfacial free energy with a hard-sphere reference potential will 
not be adequate to predict anisotropy, and that a study of the anisotropy for a variety of different 
possible repulsive potentials would be very useful. 

In recent work\cite{Laird01}, we have pointed out that the interfacal free energy for simple systems with 
face-centered cubic (fcc) crystal structures can be quantitavely described by a hard-sphere model.
It is useful to check this hypothesis here, since the LJ system is a standard molecular model that 
freezes to an fcc crystal, and since the interfacial free energy was determined in this study directly 
and not indirectly from nucleation data. The hard-sphere model predicts
\begin{equation}
\gamma_0(HS) = 0.617 kT/\sigma^2 \; .
\label{hsmodel}
\end{equation}
where $\sigma^2$ is the hard-sphere diameter. For our Lennard-Jones system we can define an effective
(temperature dependent) hard-sphere diameter using the Barker-Henderson criterion\cite{Hansen86} from 
liquid-state perturbation theory
\[
\sigma_{eff} = \int_0^\infty \{1 - exp[-u_r(r)/kT]\} dr \;,
\]
where $u_r(r)$ is the repulsive part of the potential, which we define in the Week-Chandler-Anderson
sense\cite{Hansen86} as the full potential truncated (and shifted) to zero beyond the minimum of 
the attractive well. This proceedure yields values of $\sigma_{eff}$ of 1.032$\sigma$, 1.016$\sigma$
and 1.000$\sigma$, for $T^* = 0.617$, 1.0 and 1.5, respectively. Inserting these values into 
Eq.~\ref{hsmodel} yields predicted values of $\gamma_0$ (in units of $\epsilon/\sigma^2$) of 
0.36, 0.60 and 0.93, for  $T^* = 0.617$, 1.0 and 1.5, respectively. The agreement with the values listed
for LJ in Table~\ref{tab:aniso} is excellent at the lower temperature, but overestimates the actual value
by several percent at the higher temperatures. This agreement gives more evidence to support the general 
hypothesis\cite{Laird01} that the interfacial free energy of close packed systems is largely determined by
packing considerations, not energy. 

It is interesting to note that the anisotropy parameters $\epsilon_1$ and $\epsilon_2$  for the truncated 
LJ potential at the triple point ($T^*$ = 0.617 and essentially zero pressure) are identical to those
calculated\cite{Asta02} for Ni at 1.00 atm (also essentially zero pressure). The fact that the parameters
are {\em exactly} the same for these two (essentially) zero pressure systems is, given the error bars,
probably coincidental; however, the data does show that the anisotropy for Ni is better modelled by a LJ
potential than by a hard-sphere potential. More study is required to determine the exact role that
details of the potential play in determining interfacial anisotropy. 

\section{Aknowledgements:}
This work was performed using the University of Leicester Mathematical 
Modelling Centre's supercomputer, which was purchased through the EPSRC strategic 
equipment initiative. In addition, BBL gratefully acknowledges support from
the National Science Foundation under Grant No. CHE9970903. 

\bibliography{master}
\bibliographystyle{laird_notitle}
\newpage

\end{document}